\documentclass[aps,a4paper,twocolumn]{revtex4-2}
\usepackage[utf8]{inputenc}
\usepackage{booktabs}
\usepackage{mathtools}
\usepackage{amsmath}
\usepackage{physics}
\usepackage{dsfont}
\usepackage[colorlinks,bookmarks=false,linkcolor=blue,urlcolor=blue,citecolor=magenta,linktocpage=true]{hyperref}
\usepackage[dvipsnames]{xcolor}
\usepackage{amssymb}
\usepackage{amsthm}
\usepackage{enumitem}
\usepackage[paperwidth=210mm,paperheight=297mm,centering,hmargin=2.0cm,vmargin=2.5cm]{geometry}
\usepackage{cancel}
\usepackage{bbm}

\DeclareMathAlphabet{\mathbbb}{U}{bbold}{m}{n}

\newcommand{\1}{\mathds{1}}
\newcommand{\bs}[1]{\boldsymbol{#1}}
\newcommand{\iu}{\mathrm{i}\mkern1mu}
\newcommand{\Mdu}[3]{{#1}_{#2}^{\phantom{#2}#3}}
\newcommand{\Mud}[3]{{#1}^{#2}_{\phantom{#2}#3}}

\begin{document}

\title{Casimirs of the Virasoro Algebra}
\author{Jean-François Fortin}
\email{jean-francois.fortin@phy.ulaval.ca}
\author{Lorenzo Quintavalle}
\email{lorenzo.quintavalle.1@ulaval.ca}
\author{Witold Skiba}
\email{witold.skiba@yale.edu}
\affiliation{$^\star$$^{\dagger}$ D\'epartement de Physique, de G\'enie Physique et d'Optique, Universit\'e Laval, Qu\'ebec, QC~G1V~0A6, Canada,\\
$^\ddagger$ Department of Physics, Yale University, New Haven, CT 06520, USA}

\begin{abstract}
We explicitly solve a recurrence relation due to Feigin and Fuchs to obtain the Casimirs of the Virasoro algebra in terms of the inverse of the Shapovalov form.  Combined with our recent result for the inverse Shapovalov form, this allows us to write the Casimir operators as linear combinations of products of singular vectors. 
\end{abstract}
\maketitle

\section{Introduction}

Casimir operators of an algebra $\mathfrak{g}$ are elements of the center of the corresponding universal enveloping algebra $U(\mathfrak{g})$.  As such, they commute with all elements of $\mathfrak{g}$.  This fundamental property makes Casimirs powerful tools in the theory of representations.  For example, since the number of independent Casimirs is equal to the rank of $\mathfrak{g}$ when it is semisimple, there exist proper sets of independent Casimirs that are in one-to-one correspondence with the set of numbers labeling irreducible representations.  In other words, for any state in an irreducible representation, acting with a proper set of Casimirs determines to which (unitary) irreducible representation that state belongs.

For some algebras, the number of Casimirs in the universal enveloping algebra is not sufficient to completely label irreducible representations.  In these cases, one can introduce generalized Casimirs, which are elements of the center of a natural extension of the universal enveloping algebra including infinite sums of products of generators.  In the context of this paper, we will not distinguish between generalized Casimirs and Casimirs, using the latter term for both.

Amongst many others throughout physics, Casimirs have several applications in conformal field theory (CFT).  For example, for finite-dimensional conformal algebras, there exist commuting differential operators associated to Casimirs~\cite{Dolan:2011dv,Buric:2021ywo} whose eigenfunctions are conformal blocks---the building blocks of correlation functions decomposed through the operator product expansion.  In two-dimensional CFTs~\cite{DiFrancesco:1997nk}, the global conformal algebra, which is isomorphic to $\mathfrak{sl}(2,\mathbb{R})\oplus\mathfrak{sl}(2,\mathbb{R})$, gets promoted to two copies of the celebrated infinite-dimensional Virasoro algebra.  Many aspects of the Virasoro representation theory have been studied extensively~\cite{Iohara:book}.  
For instance, the presence of the singular vectors~\cite{Kac:1978ge,Feigin:1981st,FeiginFuchs:book} has had important consequences, as they led to exact solutions of the so-called minimal models~\cite{Belavin:1984vu}.  It might thus be surprising that one of the two independent (generalized) Casimirs of the Virasoro algebra, which is of rank two, is yet to be fully determined.

Virasoro Casimirs were first studied in~\cite{FeiginFuchs:Casimir} from their action on modules over the Virasoro algebra.  The systematic method presented there leads to a recurrence relation that translates into level expansions for the Virasoro Casimirs.  In their present form, the results of~\cite{FeiginFuchs:Casimir} are however not suited to find all-level solutions.  Virasoro Casimirs also appeared in string field theory~\cite{Thorn:1985fa,Kaku:1985ui,Neveu:1985nd,Neveu:1985jv,Friedan:1985tu,Siegel:1985tw,Banks:1985ff,Kaku:1985sc,Awada:1985vk,Neuberger:1986ui,Kaku:1987vx}, where a particular choice of Casimirs is related to the string gauge-covariant action.  In all cases, the Virasoro Casimir associated with the string gauge-covariant action was ultimately expanded to some low orders as in~\cite{FeiginFuchs:Casimir}.

The goal of this paper is to find a complete solution for generic Virasoro Casimirs, which we will obtain with the help of a refinement of the recurrence relation due to Feigin and Fuchs~\cite{FeiginFuchs:Casimir}.  Our results are general and expressible in terms of the inverse of the Shapovalov form~\cite{Brower:1971qr,Shapovalov1972}, which we recently determined in terms of the singular vectors and their conformal dimensions~\cite{Fortin:2024xir}.  We thus provide new expressions for Virasoro Casimirs as functions of the singular vector data, although general formulae for singular vectors are only known in some simple cases~\cite{Benoit:1988aw,Millionshchikov2016}. 

This paper is organized as follows.  In Section~\ref{SVir} we present a lightning review of the Virasoro algebra to establish our notation.  Section~\ref{SKilling} focuses on Killing forms and shows that this approach is not suited for Casimirs of the Virasoro algebra.  In Section~\ref{SComm} we use the defining property of Casimirs to obtain the recurrence relation of Feigin and Fuchs~\cite{FeiginFuchs:Casimir}.  Our approach refines their recurrence relation and allows us to solve it in all generality in terms of the inverse Shapovalov form, and ultimately singular vectors.  Section~\ref{SDependence} investigates relations between the generic Casimirs computed in the previous section, while we conclude with an outlook on the implications of our results in Section~\ref{SConc}.  Finally, Appendix~\ref{AppCas} shows that our results extend to arbitrary algebras.

\section{Virasoro Algebra}\label{SVir}

The infinite-dimensional Virasoro algebra is spanned by $\{L_n\,\vert\,n\in\mathbb{Z}\}$ together with a central element $\hat{c}$.  The commutation relations are
\begin{equation}
\begin{gathered}
[L_n,L_m]=(n-m)L_{n+m}+\frac{\hat{c}}{12}n(n^2-1)\delta_{n+m,0},\\
[\hat{c},L_n]=0,
\end{gathered}\label{EqLAlg}
\end{equation}
and the global conformal algebra is the subalgebra of~\eqref{EqLAlg} spanned by $\{L_{\pm1},L_0\}$.

Irreducible representations of the Virasoro algebra are constructed from Verma modules.  A Verma module $V(h,c)$, indexed by the central charge $c$ and the conformal dimension $h$, is built from a lowest-weight state $\ket{h,c}$ and its descendants
\begin{equation}
L_{-\iu}\ket{h,c}\equiv L_{-i_{l_\iu}}\cdots L_{-i_1}\ket{h,c}.\label{EqLOket}
\end{equation}
Here the lowest-weight state satisfies
\begin{equation}
\begin{aligned}
L_0\ket{h,c}&=h\ket{h,c},\\
\hat{c}\ket{h,c}&=c\ket{h,c},\\
L_{n>0}\ket{h,c}&=0,
\end{aligned}\label{EqVM}
\end{equation}
and descendants are characterized by $\iu=(i_1,\cdots,i_{l_\iu})$ which is an ordered partition of level $|\iu|=\sum_{j=1}^{l_\iu}i_j$ and length (or number of parts) $l_\iu$.  Ordered partitions of level $\ell$ are thus elements of
\begin{equation}
\mathbb{O}_\ell=\left\{\iu=(i_1,\cdots,i_{l_\iu})\in\mathbb{Z}_{+}^{l_\iu}\,\bigg\vert\,|\iu|=\ell\right\}.\label{EqOPart}
\end{equation}

Arbitrary states can be expressed as linear combinations of~\eqref{EqLOket}, but the commutation relations~\eqref{EqLAlg} imply that they are not linearly independent.  A basis is furnished through standard partitions, as in
\begin{equation}
L_{-\mu}\ket{h,c}\equiv L_{-\mu_{l_\mu}}\cdots L_{-\mu_1}\ket{h,c},\label{EqLPket}
\end{equation}
where $\mu=(\mu_1,\cdots,\mu_{l_\mu})$ is an element of
\begin{equation}
\begin{aligned}
\mathbb{P}_\ell&=\biggl\{\mu=(\mu_1,\cdots,\mu_{l_\mu})\in\mathbb{Z}_{+}^{l_\mu}\,\bigg\vert\,|\mu|=\ell\\
&\qquad\And1\leq\mu_1\leq\cdots\leq\mu_{l_\mu}\biggr\}.
\end{aligned}\label{EqPPart}
\end{equation}
Hence, in the following we use Latin letters $\iu,\ldots$, typeset in roman, for ordered partitions and Greek letters $\mu,\ldots$ for standard partitions.  Moreover, the central charge $c$ will be parametrized by the more useful quantity $t$ through
\begin{equation}
c=13-6t-\frac{6}{t},\label{Eqct}
\end{equation}
and, to simplify notation, it will be dropped from the lowest-weight states, $\ket{h}\equiv\ket{h,c}$.

\subsection{Singular Vectors}

Verma modules are irreducible representations of the Virasoro algebra when they do not contain singular (\textit{a.k.a.}~null) vectors.  Singular vectors occur for specific lowest-weight conformal dimensions dictated by
\begin{equation}
h_{\expval{r,s}}=\frac{t(r+1)-(s+1)}{2}\frac{(r-1)-t^{-1}(s-1)}{2},\label{EqhSV}
\end{equation}
where $(r,s)$ is any pair of positive integers.  The associated singular vectors at level $\ell=rs$,
\begin{equation}
L_{\expval{r,s}}\ket{h_{\expval{r,s}}}\equiv v_{\expval{r,s}}^\mu L_{-\mu}\ket{h_{\expval{r,s}}},\label{EqLSV}
\end{equation}
behave as lowest-weight states in the sense that they are annihilated by the action of $L_{n>0}$.  The coefficients $v_{\expval{r,s}}^\mu$ in~\eqref{EqLSV} are functions of $r$, $s$ and $t$ and they are conventionally normalized such that $v_{\expval{r,s}}^{(1,\cdots,1)}=1$.

Singular vectors can be understood by means of the Shapovalov form~\cite{Brower:1971qr,Shapovalov1972}, the symmetric bilinear form
\begin{equation}
[S_{\ell}(h,c)]_{\mu\nu}=\bra{h}L_\mu L_{-\nu}\ket{h}.\label{EqS}
\end{equation}
Here the dual state $\bra{h}$ of the dual Verma module $\tilde{V}(h,c)$ is obtained from Hermitian conjugation, $\bra{h}\equiv(\ket{h})^\dagger$ with scalar product $\braket{h}=1$.  Since Hermitian conjugation naturally maps positive (negative) generators to negative (positive) generators as $(L_n)^\dagger\equiv L_{-n}$, a basis for the dual Verma module is thus spanned by
\begin{equation}
\bra{h}L_\mu\equiv\bra{h}L_{\mu_1}\cdots L_{\mu_{l_\mu}},
\end{equation}
with $\mu\in\mathbb{P}$.  The Shapovalov form is therefore constructed from the scalar product $\braket{\cdot}$.  Evidently, from~\eqref{EqVM} one has $\braket{v}{w}=0$ when the levels of $\ket{v}$ and $\ket{w}$ are different, thus the Shapovalov form~\eqref{EqS} is zero if $|\mu|\neq|\nu|$.  Moreover, for fixed identical levels $\abs{\mu}=\abs{\nu}\equiv\ell$, the Shapovalov form can be represented by a square matrix of dimension $p(\ell)=\dim\mathbb{P}_\ell$, which is the number of standard partitions of $\ell$.

Singular vectors at level $\ell$ occur when the rank of the matrix associated to the Shapovalov form~\eqref{EqS} is smaller than $p(\ell)$.  They can be studied quite simply from the determinant of the Shapovalov form, which at level $\ell$ is given by the Kac determinant formula
\begin{equation}
\det S_\ell(h,c)\propto\prod_{\substack{r,s\geq1\\rs\leq\ell}}(h-h_{\expval{r,s}})^{p(\ell-rs)},\label{EqKac}
\end{equation}
where the constant of proportionality, which does not depend on $h$, is unimportant to the current discussion.  Singular vectors correspond thus to eigenvectors of the Shapovalov form with vanishing eigenvalues, $[S_{rs}(h_{\expval{r,s}},c)]_{\mu\nu}v_{\expval{r,s}}^\nu=0$.  The latter identity implies in fact the presence of a singular vector in the Verma module $V(h_{\expval{r,s}},c)$ at level $rs$ given by~\eqref{EqLSV}.  As such, singular vectors are orthogonal to all the other states and must be removed from modules to generate irreducible representations.  As a final remark on singular vectors, we note that concrete expressions exist for $r$ or $s$ equal to one or two~\cite{Benoit:1988aw,Millionshchikov2016} while for arbitrary $r$ and $s$ only some properties~\cite{FeiginFuchs:book} and computational methods~\cite{Bauer:1991ai,Kent:1991qj} are known.

\subsection{Casimirs}

To achieve our goal of determining the Virasoro Casimirs, which we will do in terms of the inverse of the Shapovalov form~\eqref{EqS}, we need to introduce two important quantities, $\Mdu{S}{\mu\nu}{\alpha\beta}(L_0,\hat{c})$ and $\Mdu{R}{\mu\nu}{\lambda}$, defined by
\begin{equation}
\begin{gathered}
L_\mu L_{-\nu}=L_{-\alpha}\Mdu{S}{\mu\nu}{\alpha\beta}(L_0,\hat{c})L_\beta \, ,\\
L_{-\mu}L_{-\nu}=\Mdu{R}{\mu\nu}{\lambda}L_{-\lambda}\, ,\qquad L_\mu L_\nu=\Mdu{R}{\nu\mu}{\lambda}L_\lambda\, .
\end{gathered}\label{EqSR}
\end{equation}
In the first definition the sums are understood to be bounded by $|\alpha|\leq|\nu|$ and $|\beta|\leq|\mu|$ with $|\alpha|-|\beta|=|\nu|-|\mu|$ due to the Virasoro algebra, while in the second and third definitions the sums are constrained by $|\lambda|=|\mu|+|\nu|$.  It is interesting to point out that $\Mdu{S}{\mu\nu}{\alpha\beta}(L_0,\hat{c})$ in~\eqref{EqSR} can be seen as a generalization of the Shapovalov form~\eqref{EqS} since $\Mdu{S}{\mu\nu}{()()}(L_0,\hat{c})=[S_\ell(L_0,\hat{c})]_{\mu\nu}$ where $|\mu|=|\nu|=\ell$.

These quantities are very useful when expressing the normal ordering of Virasoro generators, and will be used repeatedly throughout our derivation of arbitrary Casimirs, which we will denote as
\begin{equation}
\mathcal{C}(M_0)=\sum_{\ell\geq0}L_{-\mu}[M_\ell(L_0,\hat{c})]^{\mu\nu}L_\nu.\label{EqLCasL0}
\end{equation}
The overall form of~\eqref{EqLCasL0} is dictated by commutativity, with $|\mu|=|\nu|=\ell$ originating from commutation with $L_0$ and $M_{\ell>0}(L_0,\hat{c})$ functions of $M_0(L_0,\hat{c})$ arising from commutation with $L_{n\neq0}$.  As stressed in the introduction, strictly speaking, \eqref{EqLCasL0} are standard Casimirs in the universal enveloping algebra only when the sum truncates, while they live in a natural extension of the universal enveloping algebra when it does not.  The latter will be the relevant case for the Virasoro algebra.  Finally, in the following we will abuse notation and sometimes designate $L_{\expval{r,s}}$ singular vectors, even though the state is missing.

\section{Casimirs from the Killing Form}\label{SKilling}

Casimir operators are often determined starting from Killing forms, in other words from the structure constants $\Mdu{f}{ij}{k}$ of the algebra $[X_i,X_j]=\Mdu{f}{ij}{k}X_k$ with elements $\{X_i\}$.  In this section we will follow this path for the Virasoro algebra, with its non-trivial structure constants $\Mdu{f}{nm}{p}=(n-m)\delta_{n+m,p}$ and $\Mdu{f}{nm}{\hat{c}}=\frac{1}{12}n(n^2-1)\delta_{n+m,0}$.

We will argue that this procedure succeeds for the global subalgebra but fails for the whole Virasoro algebra.  This failure is due to Cartan's criterion that states that the quadratic Killing form is non-degenerate if and only if the algebra is semisimple, which is not the case for Virasoro.

\subsection{Global}

The quadratic Casimir $\mathcal{Q}$ of the global algebra can be simply computed from the quadratic Killing form
\begin{equation}
g_{nm}=\Mdu{f}{ni}{j}\Mdu{f}{mj}{i}=-2(3n^2-1)\delta_{n+m,0},
\end{equation}
where the sums have support for $\{\pm1,0\}$, leading to
\begin{equation}
\begin{aligned}
\mathcal{Q}&=2g^{nm}L_nL_m=L_0^2-\frac{1}{2}L_{-1}L_1-\frac{1}{2}L_1L_{-1}\\
&=\mathcal{C}(L_0(L_0-1)),
\end{aligned}\label{EqGCas2}
\end{equation}
in the notation of~\eqref{EqLCasL0}.  Hence $M_0(L_0)=L_0(L_0-1)$, $M_1(L_0)=-1$ and the series truncates.  Having only one independent Casimir, all other Casimirs constructed from generalized Killing forms are dependent on $\mathcal{Q}$.

\subsection{Local}

Perhaps surprisingly, the quadratic Killing form for the Virasoro algebra vanishes.  Indeed, using $\zeta$-function regularization it is given by
\begin{equation}
\begin{aligned}
g_{nm}&=\Mdu{f}{ni}{j}\Mdu{f}{mj}{i}=-\sum_{j\in\mathbb{Z}}(2n-j)(n+j)\delta_{n+m,0}\\
&=[2\zeta(-2)-2n^2(1+2\zeta(0))]\delta_{n+m,0},
\end{aligned}
\end{equation}
which is zero after evaluating the $\zeta$-functions.  This degeneracy translates to all Killing forms, hence it is impossible to construct Casimirs for the Virasoro algebra from the Killing procedure.  This fact is in agreement with Cartan's criterion, as stated above.

\section{Casimirs from Commutativity}\label{SComm}

Nevertheless, it is possible to study generic Casimirs $\mathcal{C}(M_0)$ as represented in~\eqref{EqLCasL0} through their commutation properties.  Indeed, for any partition $\lambda$, commutativity leads to
\begin{equation}
\begin{aligned}
0&=[\mathcal{C}(M_0),L_{-\lambda}]\\
&=\sum_{\ell\geq0}L_{-\alpha}\left\{\Mdu{R}{\mu\rho}{\alpha}[M_\ell(L_0+|\rho|,\hat{c})]^{\mu\nu}\Mdu{S}{\nu\lambda}{\rho\beta}(L_0,\hat{c})\right.\\
&\phantom{=}\qquad\left.-\Mdu{R}{\lambda\mu}{\alpha}[M_\ell(L_0,\hat{c})]^{\mu\beta}\right\}L_\beta,
\end{aligned}
\end{equation}
where we performed a normal ordering in the two terms of the commutator through the use of~\eqref{EqSR}.  This implies
\begin{equation}
\begin{aligned}
&\Mdu{R}{\lambda\mu}{\alpha}[M_\ell(L_0,\hat{c})]^{\mu\beta}\\
&\qquad=\Mdu{R}{\mu\rho}{\alpha}[M_\ell(L_0+|\rho|,\hat{c})]^{\mu\nu}\Mdu{S}{\nu\lambda}{\rho\beta}(L_0,\hat{c}),
\end{aligned}\label{EqRMRR}
\end{equation}
for all partitions $\alpha$, $\beta$ and $\lambda$.  Since $\Mdu{R}{\lambda()}{\alpha}=\delta_\lambda^\alpha$, setting $\beta$ as the empty partition in~\eqref{EqRMRR} leads to
\begin{equation}
M_0(L_0,\hat{c})\delta_\lambda^\alpha=\Mdu{R}{\mu\rho}{\alpha}[M_{|\mu|}(L_0+|\rho|,\hat{c})]^{\mu\nu}\Mdu{S}{\nu\lambda}{\rho()}(L_0,\hat{c}),\label{EqRMRRempty}
\end{equation}
where now $|\alpha|=|\lambda|=\ell$, which can be recast as a recurrence relation that determines $M_{\ell>0}(L_0,\hat{c})$ as functions of $M_0(L_0,\hat{c})$.

To proceed, it is convenient to reintroduce explicit sums in~\eqref{EqRMRRempty} to extract the empty partition from the sum over $\rho$ with $|\rho|=\ell-|\nu|$,
\begin{equation}
\begin{aligned}
&M_0(L_0,\hat{c})\delta_\lambda^\alpha=\sum_\nu[M_\ell(L_0,\hat{c})]^{\alpha\nu}[S_\ell(L_0,\hat{c})]_{\nu\lambda}\\
&+\sum_{\substack{\mu,\nu,\rho\\0\leq|\mu|=|\nu|<\ell}}\Mdu{R}{\mu\rho}{\alpha}[M_{|\mu|}(L_0+|\rho|,\hat{c})]^{\mu\nu}\Mdu{S}{\nu\lambda}{\rho()}(L_0,\hat{c}),
\end{aligned}\label{EqMRR}
\end{equation}
and re-express~\eqref{EqMRR} in matrix notation as
\begin{equation}
M_\ell(L_0,\hat{c})=[M_0(L_0,\hat{c})\1_\ell-A_\ell(L_0,\hat{c})]S_\ell^{-1}(L_0,\hat{c}),\label{EqMFF}
\end{equation}
with the explicit definition of the $A$-matrix being
\begin{equation}
\begin{aligned}
&\Mud{[A_\ell(L_0,\hat{c})]}{\alpha}{\lambda}\\
&=\sum_{\substack{\mu,\nu,\rho\\0\leq|\mu|=|\nu|<\ell}}\Mdu{R}{\mu\rho}{\alpha}[M_{|\mu|}(L_0+|\rho|,\hat{c})]^{\mu\nu}\Mdu{S}{\nu\lambda}{\rho()}(L_0,\hat{c}).
\end{aligned}\label{EqA}
\end{equation}
Equation~\eqref{EqMFF} is the recurrence relation we were looking for, determining $M_\ell(L_0,\hat{c})$ in terms of $M_{i<\ell}(L_0,\hat{c})$ through $A_\ell(L_0,\hat{c})$, although~\eqref{EqA} is not in its most convenient form yet.

It is desirable to express~\eqref{EqA} solely in terms of the Shapovalov form and its inverse by computing the following quantity in two different ways:
\begin{equation}
\begin{aligned}
\bra{h}L_\mu L_\nu L_{-\lambda}\ket{h}&=[S_{|\mu|}(h,c)]_{\mu\rho}\Mdu{S}{\nu\lambda}{\rho()}(h,c),\\
\bra{h}L_\mu L_\nu L_{-\lambda}\ket{h}&=\Mdu{R}{\nu\mu}{\beta}[S_{|\mu|+|\nu|}(h,c)]_{\beta\lambda},
\end{aligned}
\end{equation}
where both identities originate from~\eqref{EqSR} used in different orders.  As a consequence,
\begin{equation}
\Mdu{S}{\nu\lambda}{\rho()}(L_0,\hat{c})=[S_{|\mu|}^{-1}(L_0,\hat{c})]^{\rho\mu}\Mdu{R}{\nu\mu}{\beta}[S_{|\mu|+|\nu|}(L_0,\hat{c})]_{\beta\lambda},
\end{equation}
and~\eqref{EqA} becomes
\begin{equation}
\begin{aligned}
&\Mud{[A_\ell(L_0,\hat{c})]}{\alpha}{\lambda}=\sum_{\substack{\mu,\nu,\rho,\beta,\gamma\\0\leq|\mu|=|\nu|=i<\ell}}\!\!\!\Mdu{R}{\mu\rho}{\alpha}[M_i(L_0+\ell-i,\hat{c})]^{\mu\nu}\\
&\qquad\times[S_{\ell-i}^{-1}(L_0,\hat{c})]^{\rho\gamma}\Mdu{R}{\nu\gamma}{\beta}[S_\ell(L_0,\hat{c})]_{\beta\lambda}.
\end{aligned}
\end{equation}

Going back to the recurrence relation~\eqref{EqMFF} with the $A$-matrix expressed in terms of the inverse Shapovalov form, contributions to generic Casimirs at level $\ell$ must satisfy
\begin{equation}
\begin{aligned}
&L_{-\mu}[M_\ell(L_0,\hat{c})]^{\mu\nu}L_\nu\\
&\qquad=L_{-\mu}M_0(L_0,\hat{c})[S_\ell^{-1}(L_0,\hat{c})]^{\mu\nu}L_{\nu}\\
&\qquad-\sum_{i<\ell}L_{-\alpha}L_{-\mu}[M_i(L_0+\ell-i,\hat{c})]^{\alpha\beta}\\
&\qquad\times[S_{\ell-i}^{-1}(L_0,\hat{c})]^{\mu\nu}L_{\nu}L_{\beta}.
\end{aligned}\label{EqCl}
\end{equation}
At this point, the recursive use of the improved recurrence relation~\eqref{EqCl} results in the generic Casimirs
\begin{equation}
\begin{aligned}
&\mathcal{C}(M_0)=M_0(L_0,\hat{c})+\sum_{\ell\geq1}\sum_\iu(-1)^{l_\iu-1}L_{-\mu_{l_\iu}}\cdots L_{-\mu_1}\\
&\times\{M_0(L_0+\xi_{l_\iu},\hat{c})-M_0(L_0+\ell,\hat{c})\}\\
&\times[S_{i_{l_\iu}}^{-1}(L_0+\xi_{l_\iu},\hat{c})]^{\mu_{l_\iu}\nu_{l_\iu}}\cdots[S_{i_1}^{-1}(L_0+\xi_1,\hat{c})]^{\mu_1\nu_1}\\
&\times L_{\nu_1}\cdots L_{\nu_{l_\iu}},
\end{aligned}\label{EqSoln}
\end{equation}
where $\iu=(i_1,\cdots,i_{l_i})\in\mathbb{O}_\ell$ are ordered partitions of length $l_\iu$ such that $|\iu|=\ell$.  Here the $\mu$'s and $\nu$'s are standard partitions (not parts thereof) such that $|\mu_k|=|\nu_k|=i_k$, thus $\mu_k=(\mu_{k1},\cdots,\mu_{kl_{\mu_k}})\in\mathbb{P}_{i_k}$ and similarly for $\nu$'s.  The remaining quantity appearing in~\eqref{EqSoln} is defined for an ordered partition $\iu$ as
\begin{equation}
\xi_j=\sum_{1\leq k<j}i_k,\qquad\xi_1\equiv0.\label{Eqs}
\end{equation}
Equations~\eqref{EqSoln} and~\eqref{Eqs} are our first main results.  We note that since $M_0(L_0,\hat{c})$ depends on both $L_0$ and $\hat{c}$, setting $M_0(L_0,\hat{c})=M_0(L_0)$ and $M_0(L_0,\hat{c})=M_0(\hat{c})$ provides two (families of) independent Casimirs, as expected.  The central charge Casimirs are however trivial, in the sense that $\mathcal{C}(M_0(\hat{c}))=M_0(\hat{c})$, hence for what follows they will be omitted from the discussion.  Furthermore, our results generalize directly to more complicated algebras, as discussed in Appendix~\ref{AppCas}.

It is now straightforward to apply \eqref{EqSoln} and \eqref{Eqs} to both the global subalgebra and the Virasoro algebra.

\subsection{Global}

Generic global Casimirs simplify greatly since there is only one partition per level, given by $\mu=(1,\cdots,1)$, and the inverse Shapovalov form is $S_\ell^{-1}(L_0)=\frac{1}{\ell!(2L_0)_\ell}$.  Using~\eqref{EqSoln} and~\eqref{Eqs}, they are thus
\begin{equation}
\mathcal{C}^g(M_0)=\sum_{\ell\geq0}L_{-1}^\ell M_\ell(L_0)L_1^\ell,\label{EqSolnG}
\end{equation}
with
\begin{equation}
\begin{aligned}
M_\ell(L_0)&=\sum_{\substack{\iu\\|\iu|=\ell}}\frac{(-1)^{l_\iu-1}\{M_0(L_0+\xi_{l_\iu})-M_0(L_0+\ell)\}}{i_{l_\iu}!(2L_0+2\xi_{l_\iu})_{i_{l_\iu}}\cdots i_1!(2L_0+2\xi_1)_{i_1}}\\
&=\sum_{k\geq0}\frac{(-\ell)_k(2L_0+2k-1)}{(2L_0+k-1)_{\ell+1}\ell!k!}M_0(L_0+k),
\end{aligned}\label{EqGCasMl}
\end{equation}
for any function $M_0(L_0)$.

As a check of our formula, setting $M_0(L_0)=L_0(L_0-1)$ in~\eqref{EqGCasMl} leads to
\begin{equation}
\begin{aligned}
M_\ell(L_0)&=-\frac{2(-\ell+2)_\ell(L_0+\ell)}{(2L_0+\ell)_{\ell+1}(\ell!)^2}\\
&\times[(\ell-1)L_0(L_0-1)+\ell^2(2L_0+\ell)],
\end{aligned}
\end{equation}
which reproduces the quadratic Casimir $\mathcal{Q}$~\eqref{EqGCas2}.

\subsection{Local}

It was shown in~\cite{Fortin:2024xir} that the inverse Shapovalov form for the Virasoro algebra can be expressed in terms of products of singular vectors as
\begin{equation}
\begin{aligned}
&L_{-\mu}[S_\ell^{-1}(L_0,\hat{c})]^{\mu\nu}L_\nu=\sum_{\bs{r}\cdot\bs{s}=\ell}L_{\expval{r_m,s_m}}\cdots L_{\expval{r_1,s_1}}\\
&\times\frac{q_{\expval{r_1,s_1},\cdots,\expval{r_m,s_m}}}{L_0-h_{\expval{r_1,s_1}}}L_{\expval{r_1,s_1}}^\dagger\cdots L_{\expval{r_m,s_m}}^\dagger,
\end{aligned}\label{EqInvS}
\end{equation}
where the coefficients are given by
\begin{equation}
\begin{aligned}
&q_{\expval{r_1,s_1},\cdots,\expval{r_m,s_m}}\\
&\quad=\frac{\prod_{i=1}^mq_{\expval{r_i,s_i}}}{\prod_{j=2}^m\left(h_{\expval{r_{j-1},s_{j-1}}}+r_{j-1}s_{j-1}-h_{\expval{r_j,s_j}}\right)},
\end{aligned}
\end{equation}
and
\begin{equation}
\frac{1}{q_{\expval{r,s}}}\!=\!\frac{2(-1)^{rs+1}}{rs}\!\prod_{j=1}^r(jt)_s(-jt)_s\!\prod_{k=1}^s(k/t)_r(-k/t)_r.
\end{equation}
Plugging~\eqref{EqInvS} into~\eqref{EqSoln} we obtain an explicit form for generic Virasoro Casimirs in terms of the singular vectors and their conformal dimensions.  Indeed,
\begin{widetext}
\begin{equation}
\begin{aligned}
\mathcal{C}(M_0)&=M_0(L_0,\hat{c})+\sum_{\ell\geq1}\sum_\iu(-1)^{l_\iu-1}\sum_{\bs{r}^j\cdot\bs{s}^j=i_j}\left[\prod_{j=l_\iu}^1L_{\expval{r_{m_j}^j,s_{m_j}^j}}\cdots L_{\expval{r_1^j,s_1^j}}\right]\\
&\times\{M_0(L_0+\xi_{l_\iu},\hat{c})-M_0(L_0+\ell,\hat{c})\}\left[\prod_{j=l_\iu}^1\frac{q_{\expval{r_1^j,s_1^j},\cdots,\expval{r_{m_j}^j,s_{m_j}^j}}}{L_0+\xi_j-h_{\expval{r_1^j,s_1^j}}}\right]\left[\prod_{j=1}^{l_\iu}L_{\expval{r_1^j,s_1^j}}^\dagger\cdots L_{\expval{r_{m_j}^j,s_{m_j}^j}}^\dagger\right],
\end{aligned}\label{EqSolnL}
\end{equation}
\end{widetext}
which is our last main result.  Here it is understood that $\prod_{j=l_\iu}^1a_j\equiv a_{l_\iu}\cdots a_1$, contrary to $\prod_{j=1}^{l_\iu}a_j\equiv a_1\cdots a_{l_\iu}$.

We note in passing that~\eqref{EqMFF} is the recurrence relation presented in~\cite{FeiginFuchs:Casimir}, although in~\cite{FeiginFuchs:Casimir} the $A$-matrix was left unspecified and, as a direct consequence, the recurrence relation could not be solved.

\section{(In)Dependence}\label{SDependence}

Ignoring the trivial case of the central charge, there exists a single non-trivial independent Casimir for either the Virasoro algebra or its global subalgebra.  Therefore, generic Casimirs obtained in~\eqref{EqSoln}, which depend on the arbitrary function $M_0(L_0,\hat{c})$, must be related.

\subsection{Global}

The simplicity of the quadratic global Casimir $\mathcal{Q}$, obtained through the Killing form in~\eqref{EqGCas2}, renders it a natural candidate from which all possible global Casimirs should be expressible.  It is actually straightforward to argue that generic global Casimirs can be written as
\begin{equation}
\mathcal{C}^g(M_0)=M_0\left(\frac{1}{2}\left(1-\sqrt{1+4\mathcal{Q}}\right)\right).\label{EqGCasGCas2}
\end{equation}
Indeed, \eqref{EqGCasGCas2} can be easily understood by formally expanding the square root in terms of powers of the quadratic global Casimir according to
\begin{equation}
\begin{aligned}
\frac{1}{2}\left(1-\sqrt{1+4\mathcal{Q}}\right)&=-\sum_{k\geq1}2^{2k-1}\genfrac{(}{)}{0pt}{}{1/2}{k}\mathcal{Q}^k\\
&=L_0+\text{terms with $L_{n\neq0}$},
\end{aligned}
\end{equation}
which generates $L_0$, followed by a second formal expansion of $M_0$ around $L_0$ resulting in $\mathcal{C}^g(M_0)=M_0(L_0)+\cdots$, in agreement with~\eqref{EqLCasL0}.

\subsection{Local}

Since the Killing procedure failed, there is \textit{a priori} no obvious natural choice for the independent Casimir for the Virasoro algebra.  To make progress, it is enlightening to first introduce the Fock space representation of the local algebra~\eqref{EqLAlg}.  It is constructed from the infinite Heisenberg algebra $\{a_n|n\in\mathbb{Z}\}$ with commutation relations
\begin{equation}
[a_n,a_m]=n\delta_{n+m,0},\label{EqFAlg}
\end{equation}
which allow representing the Virasoro generators as
\begin{equation}
L_n=\frac{1}{2}\sum_{m\in\mathbb{Z}}:a_{n-m}a_m:-\frac{1}{\sqrt{2}}\left(\sqrt{t}-\frac{1}{\sqrt{t}}\right)(n+1)a_n.\label{EqLnan}
\end{equation}
The normal ordering in~\eqref{EqLnan}, defined as
\begin{equation}
:a_na_m:\,=\begin{cases}a_na_m&m\geq n\\a_ma_n&m<n\end{cases},
\end{equation}
leads to
\begin{equation}
[L_n,a_m]=-ma_{n+m}-\frac{1}{\sqrt{2}}\left(\sqrt{t}-\frac{1}{\sqrt{t}}\right)n(n+1)\delta_{n+m,0},\label{EqLnam}
\end{equation}
and ultimately implies the Virasoro algebra~\eqref{EqLAlg} with central charge corresponding to~\eqref{Eqct}.

By direct translation of~\eqref{EqLCasL0}, local Casimirs in Fock space take the form $\mathcal{C}^F(m_0)=\sum_{\ell\geq0}a_{-\mu}[m_\ell(a_0,\hat{c})]^{\mu\nu}a_\nu$ with the commutativity conditions $[\mathcal{C}^F(m_0),a_n]=0$ for all $n\in\mathbb{Z}$.  However, the Heisenberg algebra~\eqref{EqFAlg} implies that local Casimirs are functions of $a_0$ only.  Hence, following the global case, local Casimirs~\eqref{EqLCasL0} are given by
\begin{equation}
\begin{aligned}
\mathcal{C}(M_0)&=M_0\left(\frac{1}{2}a_0^2-\frac{1}{\sqrt{2}}\left(\sqrt{t}-\frac{1}{\sqrt{t}}\right)a_0,\hat{c}\right)\\
&=m_0(a_0,\hat{c})=\mathcal{C}^F(m_0),
\end{aligned}\label{EqLCasFCas}
\end{equation}
since formally expanding $a_0$ in the Virasoro generators gives
\begin{equation}
\frac{1}{2}a_0^2-\frac{1}{\sqrt{2}}\left(\sqrt{t}-\frac{1}{\sqrt{t}}\right)a_0=L_0+\text{terms with $L_{n\neq0}$},
\end{equation}
according to~\eqref{EqLnan}, and a second formal expansion of $M_0$ around $L_0$ generates $\mathcal{C}(M_0)=M_0(L_0,\hat{c})+\cdots$.

Therefore, the local equivalent of functions of $\frac{1}{2}\left(1-\sqrt{1+4\mathcal{Q}}\right)$ in~\eqref{EqGCasGCas2} corresponds now to functions of $\frac{1}{2}a_0^2-\frac{1}{\sqrt{2}}\left(\sqrt{t}-\frac{1}{\sqrt{t}}\right)a_0$ in~\eqref{EqLCasFCas}.

From this discussion, we can identify a few possible natural choices for which Casimir to consider:
\begin{itemize}[leftmargin=10pt]
\item the natural Casimir operator of the Heisenberg algebra given by
\begin{equation}
\hspace*{10pt}a_0=\mathcal{C}\!\left(\!\frac{1}{\sqrt{2}}\!\left(\!\sqrt{t}-\!\frac{1}{\sqrt{t}}\right)\pm\sqrt{2L_0+\frac{1}{2}\!\left(t-2+\frac{1}{t}\right)}\right)\!,\!
\label{Eqa}
\end{equation}
as dictated by~\eqref{EqLnan}, with the plus (minus) sign corresponding to $t\in[0,1]$ ($t\notin[0,1]$);
\item $M_0(L_0,\hat{c})=L_0$, which provides the simplest expression for all coefficients;
\item $M_0(L_0,\hat{c})=L_0(L_0-1)$, which asymptotes to the quadratic global Casimir in the $c\to\infty$ limit.
\end{itemize}

\section{Conclusions}\label{SConc}

In this work, we derived an improved version of a recurrence relation for Virasoro Casimirs due to Feigin and Fuchs~\cite{FeiginFuchs:Casimir} and solved it in terms of the inverse of the Shapovalov form.  Our result, displayed in~\eqref{EqSoln}, yields an all-level solution for generic Virasoro Casimirs in terms of an arbitrary function of the Virasoro generator $L_0$.  In our previous work~\cite{Fortin:2024xir}, we determined the inverse Shapovalov form in terms of the singular vector data. Combining this with our result~\eqref{EqSoln}, we have thus succeeded in expressing generic Virasoro Casimirs as functions of the singular vectors and their conformal dimensions~\eqref{EqSolnL}.

Several interesting consequences arise as byproducts of our work.  First, our result delineates the pole structure of generic Virasoro Casimirs, which are located at the singular vector conformal dimensions and integral shifts thereof.  Moreover, any combination of generators $L_{n\neq0}$ appearing in generic Virasoro Casimirs can be grouped into products of singular vectors $L_{\expval{r,s}}$.

A second interesting consequence is an expression for the element $a_0$ of the infinite Heisenberg algebra in terms of Virasoro generators.  Indeed, since $a_0$ is a Virasoro Casimir, it has an expansion of the type $a_0=\mathcal{C}(M_0)$ for some appropriate choice of $M_0(L_0,\hat{c})$ given in~\eqref{Eqa}.  In other words, our result leads to an equation effectively inverting the relation~\eqref{EqLnan} for $a_0$.  An intriguing future avenue of research would be to invert~\eqref{EqLnan} for all $a_n$, that is expressing all elements of the infinite Heisenberg algebra in terms of the elements of the Virasoro algebra.

As mentioned in the introduction, the gauge-covariant action of strings is related to a particular Virasoro Casimir, thus our formulation in terms of singular vectors might have potential applications in that context.

Yet another interesting outcome of our work is an expression for generic Casimirs of arbitrary algebras, as discussed in Appendix~\ref{AppCas}.  This result can potentially be leveraged in the context of other infinite-dimensional algebras, like the Kac-Moody algebras.

Finally, our all-level expression of the generic Virasoro Casimirs might be of interest in the study of Virasoro conformal blocks.  For example, for four-point correlation functions, expressing the Virasoro generators in our Casimirs in terms of differential operators acting on two points, it is possible to transform our Casimirs into differential operators of the cross-ratio.  These Casimir differential operators would have as eigenfunctions conformal blocks for exchanged states $\ket{h,c}$ with eigenvalues $M_0(h,c)$.  Obviously, progress in this direction requires knowing the explicit form of the singular vectors.  Moreover, the resulting Casimir differential operators would be of infinite order.  Nevertheless, such objects might shed some light on the structure of Virasoro conformal blocks, which are notoriously complicated.  We aim to address these questions in the future.

\begin{acknowledgments}
We thank Micheal Lau, Christopher Raymond, David Ridout, and especially Pierre Mathieu for illuminating discussions.  JFF is grateful to the CERN theory group for its hospitality.  This work was supported by NSERC (JFF and LQ) and the US Department of Energy under grant DE-SC00-17660 (WS).
\end{acknowledgments}

\appendix

\section{Casimirs for Arbitrary Algebras}\label{AppCas}

Equation~\eqref{EqSoln} is general in that it translates easily to arbitrary algebras of rank $r$ with a Cartan-Weyl basis $\{H^i,E^\alpha,E^{-\alpha}\}$ such that $i\in\{1,\cdots,r\}$ and $\alpha=(\alpha^1,\cdots,\alpha^r)\in\{\alpha_1,\alpha_2,\cdots\}$ are the positive roots.

With $\alpha$, $\alpha_1$ and $\alpha_2$ positive roots and $|\alpha|^2=\sum_{1\leq i\leq r}\alpha^i\alpha^i$, the relevant part of the algebra is given by
\begin{equation}
\begin{gathered}
[H^i,H^j]=0,\\
[H^i,E^\alpha]=\alpha^iE^\alpha,\\
[H^i,E^{-\alpha}]=-\alpha^iE^{-\alpha},\\
[E^\alpha,E^{-\alpha}]=\frac{2}{|\alpha|^2}\sum_{1\leq i\leq r}\alpha^iH^i,\\
[E^{\alpha_1},E^{\alpha_2}]=\begin{cases}N_{\alpha_1,\alpha_2}E^{\alpha_1+\alpha_2}&\text{$\alpha_1+\alpha_2$ a positive root}\\0&\text{otherwise}\end{cases},
\end{gathered}
\end{equation}
where $N_{\alpha_1,\alpha_2}$ are constants.  Hence, for a highest-weight state $\ket{\lambda}$ with weight $\lambda=(\lambda^1,\cdots,\lambda^r)$ such that
\begin{equation}
H^i\ket{\lambda}=\lambda^i\ket{\lambda},\qquad E^\alpha\ket{\lambda}=0,
\end{equation}
where $\alpha$ is a positive root, an associated basis for the descendants can be constructed with some ordering of the positive roots such as $A=(\alpha_1,\cdots,\alpha_1,\alpha_2,\cdots,\alpha_2,\cdots)$ where $\alpha_i$ appears $p_i$ times,
\begin{equation}
E^{-A}\ket{\lambda}\equiv\cdots(E^{-\alpha_2})^{p_2}(E^{-\alpha_1})^{p_1}\ket{\lambda}.
\end{equation}
Proceeding naturally with the dual space,
\begin{equation}
\bra{\lambda}E^A\equiv\bra{\lambda}(E^{\alpha_1})^{p_1}(E^{\alpha_2})^{p_2}\cdots,
\end{equation}
the associated Shapovalov form is thus simply
\begin{equation}
\begin{gathered}
[S_{|A|}(\lambda)]^{AB}=\bra{\lambda}E^AE^{-B}\ket{\lambda},\\
|A|=\sum_jp_j\alpha_j,
\end{gathered}
\end{equation}
where we stress that $|A|$ is a vector.  As a consequence,~\eqref{EqSoln} becomes
\begin{equation}
\begin{aligned}
\mathcal{C}(M_0)&=M_0(H)+\sum_{i\geq1}\sum_{\{A,B\}}(-1)^{i-1}E^{-A_i}\cdots E^{-A_1}\\
&\times\{M_0(H-\xi_i)-M_0(H-\xi_{i+1})\}\\
&\times[S_{|A_i|}^{-1}(H-\xi_i)]_{A_iB_i}\cdots[S_{|A_1|}^{-1}(H-\xi_1)]_{A_1B_1}\\
&\times E^{B_1}\cdots E^{B_i},
\end{aligned}\label{EqSolnAlg}
\end{equation}
with
\begin{equation}
\xi_j=\sum_{1\leq k<j}|A_k|.
\end{equation}
We note that the sum over $i$ in~\eqref{EqSolnAlg} counts the number of copies of the inverse Shapovalov form.  At this point, one can indeed verify that~\eqref{EqSolnAlg} for $\mathfrak{su}(3)$ generates the standard quadratic and cubic $\mathfrak{su}(3)$ Casimirs by taking appropriate choices for $M_0$.

As a final remark, being general,~\eqref{EqSolnAlg} might be of interest in the study of other algebras where the Killing form approach fails, for instance the infinite-dimensional Kac-Moody algebras.


\begin{thebibliography}{10}
\providecommand{\href}[2]{#2}
\providecommand{\arxivref}[2]{\href{http://arxiv.org/abs/#1}{#2}}
\providecommand{\doiref}[2]{\href{http://dx.doi.org/#1}{#2}}
\providecommand{\nbbstauthor}[1]{#1}
\providecommand{\nbbstjournal}[1]{\textsf{#1}}
\providecommand{\nbbsttitle}[1]{\textit{``#1''}}
\providecommand{\nbbsturl}[1]{\texttt{#1}}
\providecommand{\nbbsteprint}[1]{\texttt{#1}}
\providecommand{\nbbststyle}{\raggedright\small\parskip0pt}
\nbbststyle

\bibitem{Dolan:2011dv}
\nbbstauthor{F.~A.~Dolan and H.~Osborn},
\nbbsttitle{{Conformal Partial Waves: Further Mathematical Results}},
\nbbsteprint{\arxivref{1108.6194}{arxiv:1108.6194}}.

\bibitem{Buric:2021ywo}
\nbbstauthor{I.~Buric, S.~Lacroix, J.~A.~Mann, L.~Quintavalle and V.~Schomerus},
\nbbsttitle{{Gaudin models and multipoint conformal blocks: general theory}},
\nbbstjournal{\doiref{10.1007/JHEP10(2021)139}{JHEP~2110,~139~(2021)}},
\nbbsteprint{\arxivref{2105.00021}{arxiv:2105.00021}}.

\bibitem{DiFrancesco:1997nk}
\nbbstauthor{P.~Di~Francesco, P.~Mathieu and D.~Senechal},
\nbbsttitle{{Conformal Field Theory}},
Graduate Texts in Contemporary Physics,
Springer-Verlag (1997),
New York.

\bibitem{Iohara:book}
\nbbstauthor{K.~Iohara and Y.~Koga},
\nbbsttitle{{Representation Theory of the Virasoro Algebra}},
Springer Monographs in Mathematics,
Springer London (2011).

\bibitem{Kac:1978ge}
\nbbstauthor{V.~G.~Kac},
\nbbsttitle{Contravariant form for infinite-dimensional Lie algebras and superalgebras},
in: \nbbsttitle{Group Theoretical Methods in Physics},
pp.~441--445,
ed.: W.~Beiglb{\"o}ck, A.~B{\"o}hm and E.~Takasugi,
Springer Berlin Heidelberg (1979),
Berlin, Heidelberg.

\bibitem{Feigin:1981st}
\nbbstauthor{B.~L.~Feigin and D.~B.~Fuks},
\nbbsttitle{{Invariant skew symmetric differential operators on the line and verma modules over the Virasoro algebra}},
\nbbstjournal{\doiref{10.1007/BF01081626}{Funct.~Anal.~Appl.~16,~114~(1982)}}.

\bibitem{FeiginFuchs:book}
\nbbstauthor{B.~L.~"Feigin and D.~B.~Fuchs},
\nbbsttitle{Representations of the Virasoro algebra},
in: \nbbsttitle{Representation of Lie groups and related topics},
pp.~465–-554,
ed.: A.~M.~Vershik and D.~P.~Zhelobenko,
Gordon \& Breach,
\href{https://api.semanticscholar.org/CorpusID:116941378}{\nbbsturl{https://api.semanticscholar.org/CorpusID:116941378}}.

\bibitem{Belavin:1984vu}
\nbbstauthor{A.~A.~Belavin, A.~M.~Polyakov and A.~B.~Zamolodchikov},
\nbbsttitle{{Infinite Conformal Symmetry in Two-Dimensional Quantum Field Theory}},
\nbbstjournal{\doiref{10.1016/0550-3213(84)90052-X}{Nucl.~Phys.~B~241,~333~(1984)}}.

\bibitem{FeiginFuchs:Casimir}
\nbbstauthor{B.~Feigin and D.~Fuks},
\nbbsttitle{{Operatory Kazimira v modulyakh nad algebroi Virasoro}},
\nbbstjournal{Dokl.~Akad.~Nauk~SSSR~269,~1057~(1983)}.

\bibitem{Thorn:1985fa}
\nbbstauthor{C.~B.~Thorn},
\nbbsttitle{{Comments on Covariant Formulations of String Theories}},
\nbbstjournal{\doiref{10.1016/0370-2693(85)90866-4}{Phys.~Lett.~B~159,~107~(1985)}},
[Addendum: Phys.Lett.B 160, 430 (1985)].

\bibitem{Kaku:1985ui}
\nbbstauthor{M.~Kaku},
\nbbsttitle{{Locality in the Gauge Covariant Field Theory of Strings}},
\nbbstjournal{\doiref{10.1016/0370-2693(85)91068-8}{Phys.~Lett.~B~162,~97~(1985)}}.

\bibitem{Neveu:1985nd}
\nbbstauthor{A.~Neveu, J.~H.~Schwarz and P.~C.~West},
\nbbsttitle{{Gauge Symmetries of the Free Bosonic String Field Theory}},
\nbbstjournal{\doiref{10.1016/0370-2693(85)90029-2}{Phys.~Lett.~B~164,~51~(1985)}}.

\bibitem{Neveu:1985jv}
\nbbstauthor{A.~Neveu and P.~C.~West},
\nbbsttitle{{Gauge Symmetries of the Free Supersymmetric String Field Theories}},
\nbbstjournal{\doiref{10.1016/0370-2693(85)90691-4}{Phys.~Lett.~B~165,~63~(1985)}}.

\bibitem{Friedan:1985tu}
\nbbstauthor{D.~Friedan},
\nbbsttitle{{String field theory}},
\nbbstjournal{\doiref{10.1016/S0550-3213(86)80025-6}{Nucl.~Phys.~B~271,~540~(1986)}}.

\bibitem{Siegel:1985tw}
\nbbstauthor{W.~Siegel and B.~Zwiebach},
\nbbsttitle{{Gauge String Fields}},
\nbbstjournal{\doiref{10.1016/0550-3213(86)90030-1}{Nucl.~Phys.~B~263,~105~(1986)}}.

\bibitem{Banks:1985ff}
\nbbstauthor{T.~Banks and M.~E.~Peskin},
\nbbsttitle{{Gauge Invariance of String Fields}},
\nbbstjournal{\doiref{10.1016/0550-3213(86)90496-7}{Nucl.~Phys.~B~264,~513~(1986)}}.

\bibitem{Kaku:1985sc}
\nbbstauthor{M.~Kaku},
\nbbsttitle{{Gauge Field Theory of Covariant Strings}},
\nbbstjournal{\doiref{10.1016/0550-3213(86)90147-1}{Nucl.~Phys.~B~267,~125~(1986)}}.

\bibitem{Awada:1985vk}
\nbbstauthor{M.~A.~Awada},
\nbbsttitle{{The Gauge Covariant Formulation of Interacting Strings and Superstrings}},
\nbbstjournal{\doiref{10.1016/0370-2693(86)90211-X}{Phys.~Lett.~B~172,~32~(1986)}}.

\bibitem{Neuberger:1986ui}
\nbbstauthor{H.~Neuberger},
\nbbsttitle{{{BRS} Cohomology and Casimir Operators}},
\nbbstjournal{\doiref{10.1016/0370-2693(87)90009-8}{Phys.~Lett.~B~188,~214~(1987)}}.

\bibitem{Kaku:1987vx}
\nbbstauthor{M.~Kaku},
\nbbsttitle{{String Field Theory}},
\nbbstjournal{\doiref{10.1142/S0217751X87000028}{Int.~J.~Mod.~Phys.~A~2,~1~(1987)}}.

\bibitem{Brower:1971qr}
\nbbstauthor{R.~C.~Brower and C.~B.~Thorn},
\nbbsttitle{{Eliminating spurious states from the dual resonance model}},
\nbbstjournal{\doiref{10.1016/0550-3213(71)90452-4}{Nucl.~Phys.~B~31,~163~(1971)}}.

\bibitem{Shapovalov1972}
\nbbstauthor{N.~N.~Shapovalov},
\nbbsttitle{{On a bilinear form on the universal enveloping algebra of a complex semisimple Lie algebra}},
\nbbstjournal{\doiref{10.1007/BF01077650}{Func.~Anal.~Appl.~6,~307~(1972)}}.

\bibitem{Fortin:2024xir}
\nbbstauthor{J.-F.~Fortin, L.~Quintavalle and W.~Skiba},
\nbbsttitle{{Virasoro completeness relation and the inverse Shapovalov form}},
\nbbstjournal{\doiref{10.1103/PhysRevD.111.085010}{Phys.~Rev.~D~111,~085010~(2025)}},
\nbbsteprint{\arxivref{2409.12224}{arxiv:2409.12224}}.

\bibitem{Benoit:1988aw}
\nbbstauthor{L.~Benoit and Y.~Saint-Aubin},
\nbbsttitle{{Degenerate Conformal Field Theories and Explicit Expression for Some Null Vectors}},
\nbbstjournal{\doiref{10.1016/0370-2693(88)91352-4}{Phys.~Lett.~B~215,~517~(1988)}}.

\bibitem{Millionshchikov2016}
\nbbstauthor{D.~V.~Millionshchikov},
\nbbsttitle{{Virasoro singular vectors}},
\nbbstjournal{\doiref{10.1007/s10688-016-0149-9}{Functional~Analysis~and~Its~Applications~50,~219~(2016)}}.

\bibitem{Bauer:1991ai}
\nbbstauthor{M.~Bauer, P.~Di~Francesco, C.~Itzykson and J.~B.~Zuber},
\nbbsttitle{{Covariant differential equations and singular vectors in Virasoro representations}},
\nbbstjournal{\doiref{10.1016/0550-3213(91)90541-5}{Nucl.~Phys.~B~362,~515~(1991)}}.

\bibitem{Kent:1991qj}
\nbbstauthor{A.~Kent},
\nbbsttitle{{Singular vectors of the Virasoro algebra}},
\nbbstjournal{\doiref{10.1016/0370-2693(91)90553-3}{Phys.~Lett.~B~273,~56~(1991)}},
\nbbsteprint{\arxivref{hep-th/9204097}{hep-th/9204097}}.

\end{thebibliography}

\end{document}